\documentclass[preprint2]{aastex}

\slugcomment{ }

\shorttitle{Mapping the X-Shaped Milky Way Bulge}
\shortauthors{Saito et al.}

\begin{document}

\title{Mapping the X-Shaped Milky Way Bulge}

\author{
R. K. Saito\altaffilmark{1}, 
M. Zoccali\altaffilmark{1}, 
A. McWilliam\altaffilmark{2}
D. Minniti\altaffilmark{1,3,4}
O. A. Gonzalez\altaffilmark{5} and
V. Hill\altaffilmark{6}}

\affil{
$^1$Departamento de Astronom\'{\i}a y Astrof\'{\i}sica, Pontificia 
    Universidad Cat\'{o}lica de Chile, Vicu\~na Mackenna 4860, 
    Casilla 306, Santiago 22, Chile; rsaito@astro.puc.cl, mzoccali@astro.puc.cl,
    dante@astro.puc.cl \\
$^2$  The Observatories of the Carnegie Institute of Washington,
      813 Santa Barbara St., Pasadena, CA 91101--1292; andy@obs.carnegiescience.edu \\
$^3$ Vatican Observatory, Vatican City State V-00120, Italy \\
$^4$ European Southern Observatory, Vitacura 3107, Santiago, Chile \\
$^5$ European Southern Observatory, Karl-Schwarzschild-Strasse 2, D-85748 Garching,
Germany; ogonzale@eso.org \\
$^6$ Universit\'e de Nice Sophia Antipolis, CNRS, Observatoire de la
C\^{o}te d'Azur, B.P. 4229, 06304 Nice Cedex 4, France; vanessa.hill@oca.eu}

\begin{abstract}

We analyzed the distribution of the RC stars throughout Galactic bulge
using 2MASS data. We mapped the position of the red clump in
1~sq.~deg.  size fields within the area $|l|\le8.5^{\circ}$ and
$3.5^{\circ}\le |b| \le8.5^{\circ}$, for a total of 170~sq.~deg. The
red clump seen single in the central area splits into two components
at high Galactic longitudes in both hemispheres, produced by two
structures at different distances along the same line of sight. The
X-shape is clearly visible in the $Z$--$X$ plane for longitudes close
to $l=0^{\circ}$ axis. Crude measurements of the space densities
of RC stars in the bright and faint RC populations are consistent with
the adopted RC distances, providing further supporting evidence that
the X-structure is real, and that there is approximate front-back
symmetry in our bulge fields. We conclude that the Milky Way bulge has
an X-shaped structure within $|l|\lesssim 2^{\circ}$, seen almost edge
on with respect to the line of sight. Additional deep NIR photometry
extending into the innermost bulge regions combined with spectroscopic
data is needed in order to discriminate among the different
possibilities that can cause the observed X-shaped structure.

\end{abstract}

\keywords{Galaxy: bulge --- Galaxy: structure --- stars: distances ---
  stars: late-type}

\section{Introduction}

In the last decade consensus has been reached that the Galactic bulge
has a boxy shape in the COBE/DIRBE near infrared map
\citep{1995ApJ...445..716D} because it contains a bar. This bar is
tilted by $\sim 20-25$ degrees from the Sun-Galactic center direction,
with axis ratios 1\,:\,0.33\,:\,0.23, and a scale length of $\sim
1.5$~kpc \cite[e.g.,][and references therein]{1994ApJ...429L..73S,
  1997ApJ...477..163S, 2005MNRAS.358.1309B, 2007MNRAS.378.1064R}. Most
of these studies make use of horizontal branch red clump (RC) stars as
tracers of the bulge 3D structure, due to their narrow range in color,
brightness, and well calibrated absolute magnitude.

However, recent analysis of the RC stars across the bulge area
suggested a more complex X-shaped structure
\citep{2010IAUS..265..279M, 2010IAUS..265..271Z, 2010ApJ...721L..28N,
  2010ApJ...724.1491M}. Along to $l=0^{\circ}$ axis, both above and
below the plane, the RC appears to split in two components. Both
components become closer and finally merge at the latitude of Baade's
Window ($b=-4^{\circ}$) and below, while they open up and diverge at
higher latitudes both above and below the plane. The analysis by
\cite{2011ApJ...730..118N} is based on the RC distribution from the
OGLE-II photometry, while \cite{2010ApJ...724.1491M} analyzed the
color magnitude diagrams (CMDs) from 2MASS on a roughly symmetric
field grid across the bulge. For the field at
$(l,b)=(0^{\circ},-6^{\circ})$, in particular,
\cite{2010ApJ...724.1491M} had deeper SOFI@NTT infrared photometry
($K_{\rm s}\sim18$~mag, in comparison to $K_{\rm s}\sim14.3$~mag from
2MASS) as well as deep optical data from WFI@ESO/MPG telescope
($V\sim23.0$~mag, $I\sim22.5$~mag).

Here we complete the analysis of the RC distribution across a much
wider bulge area using 2MASS data. The RC was mapped at positive and
negative latitudes in 170 fields of 1~sq.~deg. size, within
$|l|\le8.5^{\circ}$ and $3.5^{\circ}\le |b| \le8.5^{\circ}$.  Our
results are in agreement with previous studies, and extend their
findings, showing the presence of an X-shaped structure across the
whole Galactic bulge.


\section{Observational Data}

  \begin{figure}[ht]
  \includegraphics[bb=8cm 1.5cm 19cm 10cm,angle=-90,scale=0.50]{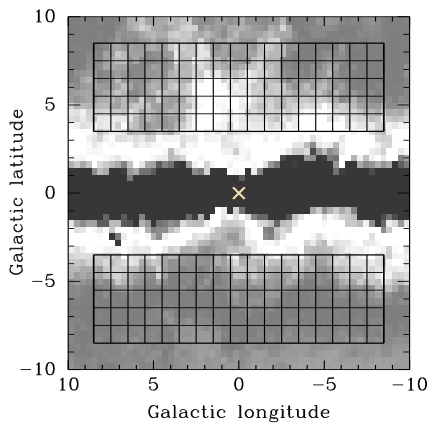}
    \caption{The bulge area analyzed here, divided in 170 square
      fields, overplotted on the \cite{1998ApJ...500..525S} dust
      map. The inner area was not used here because the 2MASS
      photometry does not reach the RC magnitude.}
  \label{area}
  \end{figure}

This analysis makes use of data from the Two Micron All Sky Survey
(2MASS) point source catalog \citep{2006AJ....131.1163S} for the
Galactic bulge region within coordinates $|l|\le8.5^{\circ}$ and
$3.5^{\circ}\le |b| \le8.5^{\circ}$. In order to map the RC position
across the bulge, the whole area was divided in square fields of
$1^\circ$ size, centered on each integer Galactic coordinate, for a
total of $17 \times 10=170$ fields (see Fig.~\ref{area}).
Unfortunately, the incompleteness of 2MASS in the region close to the
Galactic plane ($|b|<3.5^{\circ}$) becomes rather severe at the RC
magnitude, so this region has been excluded from the present analysis.

\begin{figure}[ht]
\includegraphics[bb=2.3cm 6cm 10cm 25cm,scale=0.5]{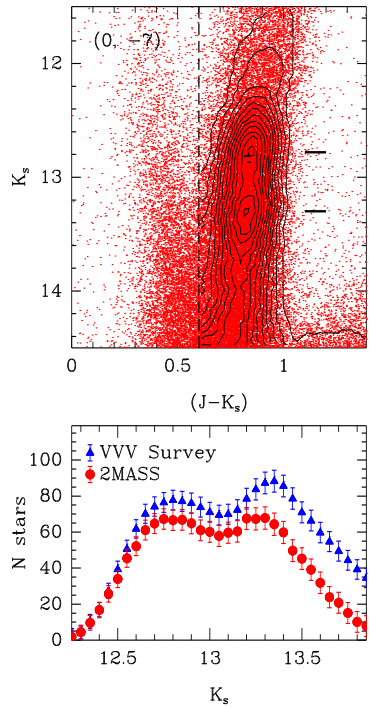}
 \caption{Top panel: Zoom up of the RCs in the 2MASS CMD of a selected
   field at $(l,b)=(0^{\circ},-7^{\circ})$. A vertical dashed line
   marks the color cut applied in order to exclude from the analysis
   the foreground disk main sequence stars on the blue side. Contours
   help the eye identify the two clumps, also marked by horizontal
   thick lines.  Bottom panel: $K_{\rm s}$ luminosity function of the
   bulge RGB near the RC, both in 2MASS and in the VVV data (see
   text).}
\label{cmds}
\end{figure}

As  discussed  by \cite{2010ApJ...724.1491M},  the most  plausible
explanation for the double RC  is the presence of two structures along
the  line of  sight. We  therefore mapped  the position  of  the RC(s)
across a large area, in order to deproject the bulge 3D structure. The
procedure was the following:

{\it  i\,}) A $K_{\rm  s}~vs.~(J-K_{\rm s})$  CMD was  constructed for
each of the 170 fields (Fig.~\ref{cmds}) and a color cut applied, fine
tuned for  each CMD in  order to separate  the bulge red  giants (RGB)
from  the  main sequence  of  disk  foreground  stars. The  luminosity
function of bulge  RGB stars was constructed in  order to identify the
RC(s).

{\it ii\,}) Using the RC(s) as distance indicator, we assume that the
two peaks seen at the bottom of Fig.~\ref{cmds} are due to the
presence of two overdensities along the line of sight. In order to map
the overdensities in 3D space, we associate a distance to each of the
two peaks, and further assume that the width of each peak is entirely
due to distance spread. This latter assumption lead us to overestimate
the distance spread, because other effects can contribute to the RC
width. The photometric errors at the magnitude of the RCs ($K_{\rm
  s}<13.5$~mag) produce uncertainties of $\Delta d\sim0.1$~kpc in the
RC width, while the differential reddening should account for $\Delta
d\sim0.2$~kpc. In addition, the RC has an intrinsic width due to the
broad metallicity distribution, and a possible spread in age and
helium content, both hard to quantify.  The assumption of the width of
the RC peaks being due only to distance allows us to directly convert
the luminosity function into a density map along the line of sight, by
assigning a distance to the $K_{\rm s}$ magnitude of each star close
to the RC.  In this way, the position of the main structures in the
map will be correct\footnote{The intrinsic width of the RC, combined
  with the fact that we did not subtract the underlying RGB luminosity
  function, could actually displace slightly the estimated position of
  the RC peak.  However, this would be negligible compared to the
  effect of a small difference in age or helium content, and in any
  case none of these effects would change our conclusion about the
  global bulge shape.}, but their width along the line of sight will
be overestimated.  The distance associated to each RC star was
calculated as:
\begin{equation}
\mu =-5+5log(d)=K-M_{\rm K}-A_{\rm K}
\end{equation}
where the absolute magnitude $M_{\rm K}$ was taken from the horizontal
branch  in  the  isochrones  of \cite{2004ApJ...612..168P}  for  solar
metallicity,   and    $A_{\rm   K}$    taken   from   the    maps   of
\cite{1998ApJ...500..525S}. This makes  the reasonable assumption that
all  the extinction happens  between us  and the  bulge, with  no dust
being present between the two bulge RCs.

  \begin{figure*}[ht]
  \includegraphics[bb=-.8cm -1cm 21cm 18cm,angle=-90,scale=0.50]{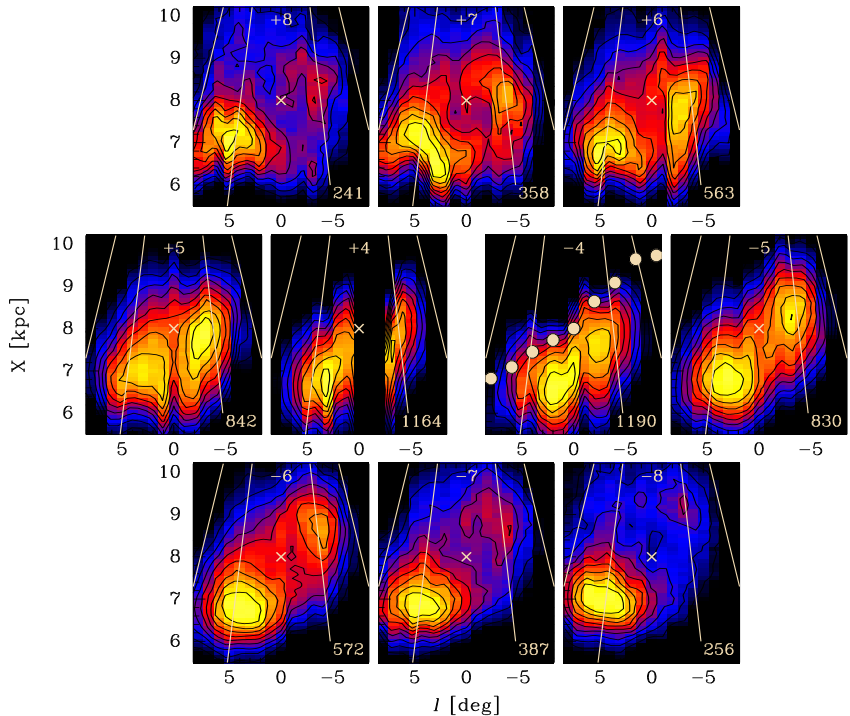}
   \caption{Density maps showing the structures traced by the RC near
     the Galactic plane, i.e., as seen from above, in slices of
     different latitudes (see labels). Individual lines of sight, at a
     given longitude, are represented by vertical strips, which are
     then merged together to form each panel. The central strip in
     each panel corresponds to $l=0^{\circ}$ with a cross marking the
     Galactic Center (assuming $R_0=8~$kpc). The panel at
     $b=-4^{\circ}$ also shows the Galactic bar as traced by
     Rattenbury et al. (2007; white dots) fitting the RC in OGLE II
     data.  The label at the bottom of each panel lists the peak value
     of the density histogram in that particular section of the 3D
     map. Contour plots may help the eye in regions of low density
     contrast. Thin white lines are lines of constant $Y$ coordinate.}
  \label{dist}
  \end{figure*}

   \begin{figure*}[ht]
  \includegraphics[bb=0cm -2.5cm 20cm 17cm,angle=-90,scale=0.55]{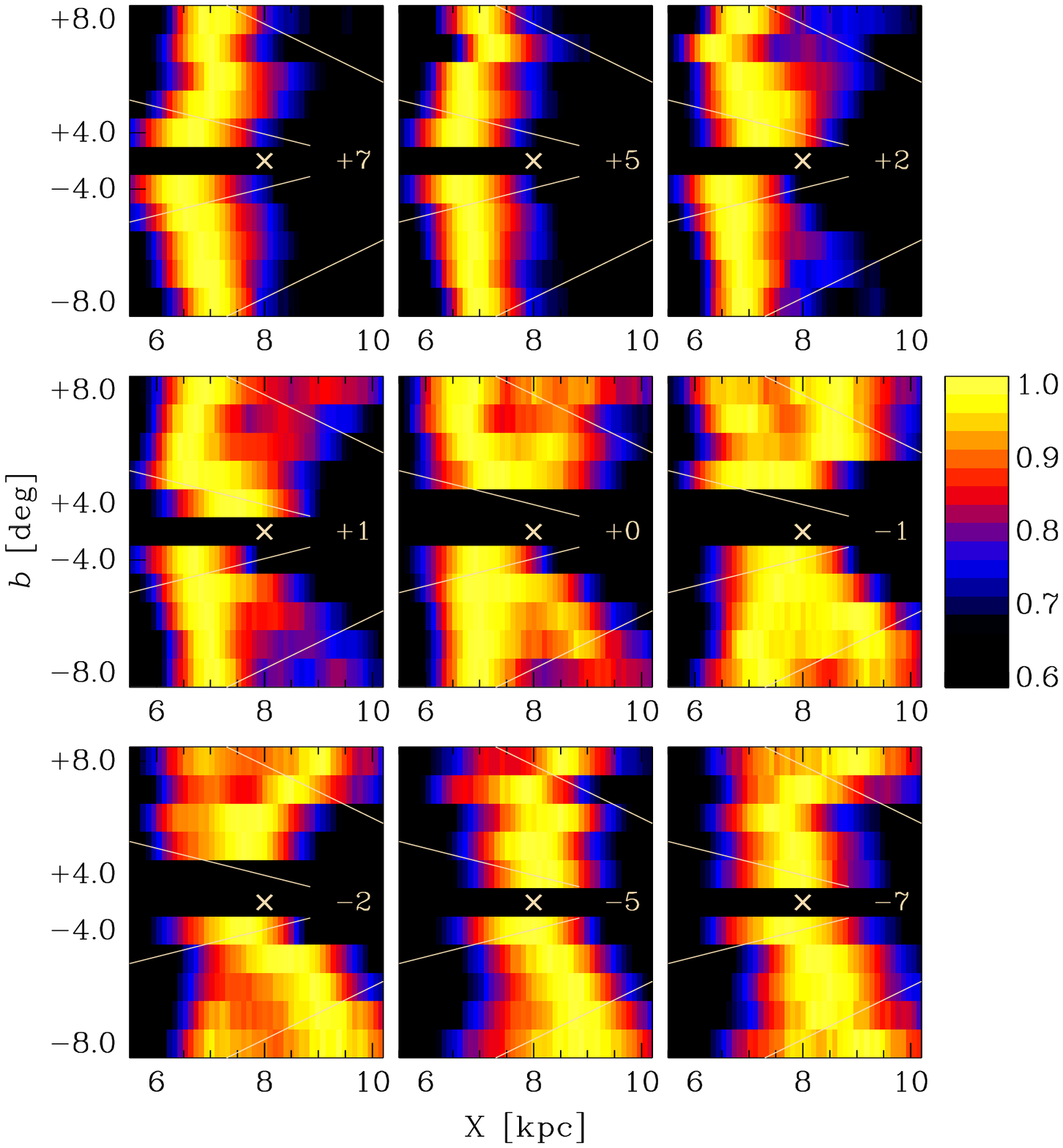}
   \caption{Density maps showing the structures traced by the RC in
     the ($X,b$) plane. Each panel corresponds to a different
     longitude (see labels), with $l=0^{\circ}$ in the central, middle
     one. A cross marks the Galactic Center (assuming $R_0=8$~kpc),
     with the Sun far in the left side, outside the figure, at
     $(X,Z)=(0,0)$. Individual lines of sight, at different latitudes,
     correspond to horizontal color strips, merged together to form
     the panels. The color scale has been normalized so that the
     histogram peak in each horizontal strip has density=1.  Note that
     the region close to the Galactic plane, for which we have no
     data, has been compressed here, and shown as a single, horizontal
     black strip. Thin white lines are lines of constant $Z$
     coordinate.  The X-shape is clearly visible for longitudes
     $|l|\leq1^{\circ}$ (middle row panels).}
   \label{xshape}
  \end{figure*}

{\it iii\,}) Distance histograms, corresponding to density maps along
the line of sight, were constructed for RC stars in each of the 170
fields, using an averaged shifted histogram, for a bin width of
$0.2$~kpc.

Figure~\ref{cmds} also shows the luminosity function of RC stars, in
the same bulge field, from the Vista Variables in the V\'ia L\'actea
ESO Public Survey (VVV; Minniti et al. 2010). For this figure, the
CASU v1.1 version of the catalog has been used\footnote{This is the
  catalog, produced by the Cambridge Astronomical Survey Unit
  (http://casu.ast.cam.ac.uk/), corresponding to the $1^{st}$ VVV data
  release, on June 2011.}.  In this field, the VVV survey reaches down
to $K_{\rm s}\sim 17.5$~mag and the photometric error at the level of
the RC is much smaller than in 2MASS.  The presence of a double RC is
fully confirmed by VVV data. The faint RC is stronger in the VVV data,
due to incompleteness of 2MASS. In fact, the discrepancy between the
two surveys increases at fainter magnitudes, due to confusion becoming
stronger and stronger in 2MASS. The VVV survey data will allow us to
map the bulge 3D structure down to the Galactic plane.  Extensive
analysis of the upper RGB and RC morphology across the bulge area
covered by VVV is ongoing and will be the subject of a dedicated paper
(Gonzalez et al.  2011, in preparation).


\section{The X-Shaped Structure}

In what follows we will discuss different sections of the bulge density
maps constructed above.

The panels in Figure~\ref{dist} show sections of the density map
roughly parallel to the Galactic plane, seen from above, each at a
given Galactic latitude.  The region within $|b|<3.5^{\circ}$ is
excluded from the analysis, so the innermost sections are at
$b=+4^{\circ}$ and $b=-4^{\circ}$, in the middle row panels. A cross
marks the Galactic Center, assuming a distance of 8~kpc \citep[e.g.,
  $R_0=8.00\pm0.6$~kpc;][]{2008ApJ...689.1044G}. The Sun would be at
$(l,X)=(0^{\circ},0)$, below and outside the limits of the plots. The
map in each panel is constructed merging 17 vertical color strips for
the fully sampled sub-panels, each showing the RC star density along a
given line of sight, at a given longitude (x-axis). Contour lines mark
equal density levels, in steps of 5$\%$ from the peak intensity. The
section at $b=+4^{\circ}$ has a vertical, black strip due to a dust
lane that prevents us to reach the RC in 2MASS, in this part. Due to
the opening of the cone of view, we see nearby things more expanded in
the sky, than far away ones. In order to show the distortion
introduced by this effect, we also show diagonal thin white lines at
constant $Y$ linear coordinate.

The color scale has been normalized in each panel so that the peak
density {\it of that section} is yellow. This stretching of the scale
allows us to trace the shape of the bulge even far away from the
plane, where stellar densities are much lower, but might give the
wrong impression that the total number of bulge stars at
$|b|=4^{\circ}$ is the same as it is at $|b|=8^{\circ}$, which is
clearly not the case.  To clarify this, the peak value of the density
histogram (corresponding to the lightest yellow) is listed at the
bottom of each panel.

Two overdensities can be seen in each panel: one closer to the Sun, at
positive Galactic longitudes, and a more distant one, at negative
longitudes. Moving away from $l=0^{\circ}$ to larger longitudes, one
of the two structures gradually disappear, leaving just the bright one
at positive longitudes, and the faint one at negative longitudes. The
two overdensities get closer to each other for sections closer to the
Galactic plane, and they almost completely merge already at
$b=-4^{\circ}$ (Baade's Window). This is the section that has been
studied extensively in the past, and the structure that one sees here
is consistent with a single, elongated overdensity, i.e., the {\it
  bar}. Clearly, only for lines of sight along to $l=0^{\circ}$
axis, i.e, in the central, vertical strip of each section, and only
for $|b|>4^{\circ}$ we intercept both overdensities, resulting in a
double RC.

Interestingly, in this map the far side of the X fades faster than
the near side, when moving away from the Galactic plane. In the next
section we will investigate whether this is a real feature or an
artifact of our data analysis. Let us concentrate, here, on the 3D
shape of the Galactic bulge in a qualitative way.

The panel at $b=-4^\circ$ also shows the Galactic bar, as traced by
\cite{2007MNRAS.378.1064R} using OGLE II data for stars at a similar
latitude. In the work of Rattenbury et al. the derived bar was
arbitrarily shifted in distance so that its center would be at 8
kpc. The angle between the structure we find and the line of sight is
clearly the same as that of the Rattebury's bar. The center of
structure, at $b=-4^\circ$, is $\sim 7$ kpc away from the Sun (see
below).

In Fig.~\ref{xshape}, each panel shows a vertical section of the
density map, parallel to $l=0^\circ$ axis, at a given longitude. The $X-b$
plane passing from $l=0^{\circ}$ is in the central, middle panel.  The
Sun would be at $(X,b)=(0,0^{\circ})$, outside each panel, on the
left. Lines of sight at different latitudes are shown as horizontal
color strips, in each panel. Again, the line of sight at
$(l,b)=(0^\circ,+4^\circ)$ is missing due to high
extinction. Thin white lines in each panel are lines of constant
height above/below the plane ($Z$ linear coordinate).

The color map has been normalized so that the peak density along each
line of sight gets the maximum yellow intensity. Again, this allows us
to trace the bulge shape even far away from the plane, but of course
the density of bulge stars at $|b|=4^{\circ}$ and $|b|=8^{\circ}$ is
not the same. Looking at $|l|\leq1^{\circ}$, the bulge splits in two
components both above and below the plane. However, when moving to
sections at positive longitudes (top panels) only the one closer to
the Sun is visible, while at negative longitudes (bottom panels) only
the far one remains.

Although we do not have data in the region close to the plane, it is
reasonable to assume that the elongated structures seen at $|b|\geq
4^\circ$ would merge in the center, completing an X-shaped
structure. If we make the exercise of joining the {\it arms} of the X
above and below the plane, we find that the whole structure is
approximately centered at a distance of $7.4\pm0.6$ kpc from the Sun,
which is consistent with the most recent determination of the distance
to the Galactic center. The latter was found to be $7.52\pm0.10$ kpc
by Nishiyama et al. (2005), $8.0\pm0.6$ kpc by Ghez et al. (2008), and
$8.33\pm0.35$ kpc by Gillessen et al. (2009). The accuracy of the
distance from the Sun to the center of the X-shaped structure found
here depends on how appropriate is the adopted value for the absolute
magnitude of the bulge RC. Uncertainties on this parameter may come
from the broad bulge metallicity distribution function, and the
unknown precise age and helium content of the bulge. However, this is
beyond the purpose of the present paper, which addresses the {\it
  shape} of the outer ($|b|>4^\circ$) Galactic bulge.

In closing this section let us comment on the effect of
incompleteness.  Incompleteness affects the present analysis in two
ways.  First, it artificially reduces the star density, especially in
the faint RC (Fig.~\ref{cmds}).  For a qualitative analysis of the
bulge shape such as the present one, this might not be a serious
bias. Second, severe incompleteness produces a cutoff in the middle of
the RC, so that the peak in the RC luminosity function looks brighter
than it really is. Indeed, when fields at $|b|<3.5^{\circ}$ were
tentatively included in the analysis, the map would show the bulk of
bulge stars getting closer and closer to the Sun, for lower and lower
$|b|$.  This effect was not trusted as real, and this is where our
decision to exclude fields at $|b|<3.5^{\circ}$ comes from.  Even for
high $|b|$, however, the extinction is highly variable on a field to
field basis, and so it is the 2MASS degree of incompleteness.  The
wiggles of the arms of the $X$ in Fig.~\ref{xshape} (see, e.g., the
map at positive latitudes, for $l=+5^{\circ}$) might in fact be due to
differential extinction and thus incompleteness.

\section{Relative density of the bright and faint RC populations}

Here we investigate the density of RC stars in the faint and bright
RCs. The RC densities not only reveal the structure and symmetry of
the X-shaped bulge, but also provide a consistency check to address
any lingering doubts that the brightness difference between the two
RCs might be due to something other than distance.

We measured the  density of RC stars as a  function of vertical height
below  the  Galactic plane,  at longitude
$l\sim+5^{\circ}$ for  the bright red clump,  and at $l\sim-3^{\circ}$
for the faint red clump; at  these latitudes the bright and faint RCs,
respectively,  are maximum \citep[see][]{2010ApJ...724.1491M}.  The RC
densities  were  measured from  30  arc-minute  radius circular  2MASS
fields,   in    $K_{\rm   s}$   frequency    histograms   similar   to
\cite{2010ApJ...724.1491M}  and  included  a linear  interpolation  to
remove  the  background   counts.   Background  subtraction  would  be
significantly improved  with the use of  isochrone fitting techniques,
but this  is beyond the  scope of the  current paper.  To  compute the
space  densities  of  the RC  stars  from  the  number counts  it  was
necessary to assume distances for the bright and faint RC populations;
we  adopted   6.4  and   9.0~kpc  distances  for   these  populations,
respectively (see McWilliam \& Zoccali 2010).

The vertical extent of our density fields ranges from $b=-6.5^{\circ}$
to $-12.6^{\circ}$ in latitude, and generally corresponds to vertical
distances, $Z$, from 1.0 to 1.4~kpc. All projection effect have been
taking into account when computing the stellar surface densities for
the bright and faint RC populations.

\begin{figure}
  \includegraphics[bb=2.3cm 6cm 10cm 25cm,scale=0.38]{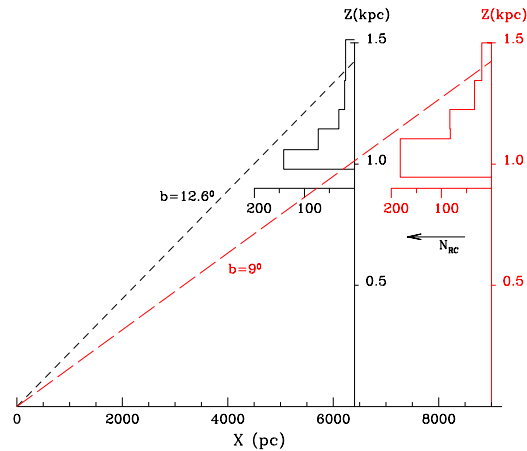}
    \caption{ Measured density of RC stars in the foreground (black)
      and background (red) RC peaks are shown; units for $N_{\rm RC}$
      are 10$^{-3}$~pc$^{-2}$.  Foreground and background RC distances
      of 6.4~kpc and 9.0~kpc were adopted from
      \cite{2010ApJ...724.1491M}, who estimated them from
      $b=-8^{\circ}$. Note that the two RC populations show very
      similar declining density trends with vertical, Z, distance
      below the plane, but are markedly different when compared with
      angular distance (i.e., latitude) from the Galactic plane.  In
      particular, at latitude $b=-9^{\circ}$ the background RC
      population is present, but barely detectable, yet the foreground
      population at the same latitude is enormous; only at
      $b=-12.6^{\circ}$ does the foreground RC density resemble the
      background at $b=-9^{\circ}$, where $Z=1.425$~kpc for both
      populations.  This supports our distance interpretation for the
      brightness difference of the two RC populations, and indicates
      that there {\em is} approximate density symmetry between the
      foreground and background bulge.  However, a slightly higher
      density of stars may be present for the background, faint, RC.}
  \label{density}
  \end{figure}

Our results for the two regions are plotted in Fig.~\ref{density},
which shows that the faint and bright RCs possess similar declining
trends in space density with distance below the Galactic plane.
However, if one considers only the angular extent below the plane
(i.e., latitude), it is clear that the faint RC density declines much
more rapidly with angle than the bright RC.  For example,
Fig.~\ref{density} shows that at $b=-9^{\circ}$ the faint RC has almost
vanished, but there are large numbers of RC stars in the bright RC at
this latitude.

Put in another way: with the assumption of approximate
foreground-background symmetry of the bulge, the trend of RC density
with latitude is consistent with a distance interpretation for the
bright and faint red clump populations.  Given the extensive
discussion of distance interpretation in \cite{2010ApJ...724.1491M},
and the consistency check discussed here, we conclude that the RC
brightness differences must, indeed, be due to distance.  Therefore,
the foreground and background bulge RC densities possess approximate
symmetry.  In detail, however, our measured densities are slightly
higher for the faint RC field.  We believe that this difference is
probably real, but could be due to errors in our analysis, such as
poor background subtraction, or in the precise distances adopted for
the two RCs.  Clearly, this small density asymmetry should be
investigated further.

\section{Discussion and Conclusions}

The position of the RC(s) was mapped across a large bulge area, using
2MASS photometry.  By using the RC(s) magnitude as distance indicator,
density maps were constructed along the line of sight, for the region
$|l|\le8.5^{\circ}$ and $3.5^{\circ}\le |b|\le8.5^{\circ}$.

The main conclusion of the present paper is that in the Milky Way
bulge has an X-shape, seen almost edge on, with an inclination of
$\sim 20^\circ$ with respect to the line of sight. The X is clearly
seen within $|l|\lesssim2^{\circ}$, with the closer arm remaining at
positive latitudes while at negative latitudes only the farther arm is
present. The {\it arms} of the X merge together both above and below
the plane, for latitudes $|b|\le 4^{\circ}$. Therefore, in Baade's
Window at $b=-4^{\circ}$ one would just see an elongated structure,
corresponding to the two lower arms being {\it almost} completely
merged.  This feature has been classically interpreted as the Galactic
bar.

Our crude measurement of the space densities of RC stars as a function
of vertical height below the Galactic plane is consistent with the
adopted RC star distances, if the bulge structures studied here are
roughly symmetric from front to back.  This removes any lingering
doubt about the interpretation of the RC $K$-band magnitudes as due to
distance, and strengthens the conclusion that the bulge contains an
X-shape structure.  Our densities are slightly higher for the
background structure; this possibility should be investigated with
more sophisticated techniques.

Several galaxies are known to host X-shaped bulges (e.g., NGC~128,
NGC~3625, NGC~4469), and some numerical simulations predict their
formation, with a particularly strong buckling of a bar
\citep{2002MNRAS.337..578P, 2005MNRAS.358.1477A,
  1995ApJ...447L..87M}. \cite{2010ApJ...724.1491M} review some of the
proposed mechanisms to form such structures.

The X-shape structure seems to be relatively concentrated, with
respect to other galaxies (e.g., NGC~4710) and probably is the
dominant component in the inner 2~kpc.  It remains to be understood
how the bulge 3D structure can be reconciled with the other features,
such as the presence of a radial metallicity gradient
\citep{2008A&A...486..177Z, 2010ApJ...725L..19B}, the kinematic
difference between the metal poor and metal rich component in Baade's
Window \citep{2010A&A...519A..77B} and the behavior of the alpha
element ratios, which seems to be constant in different fields
(Gonzalez et al. 2011). It should be kept in mind, however, that all
the properties listed above have been derived from observations on
$l\simeq0^{\circ}$ only. Just as the shape turned out to be much more
complex as soon as we analyzed a larger area, the other properties
might turn out to be a partial view of the global picture.

Radial velocities of stars in the two RCs, at $b=-8^{\circ}$ revealed
no differences \citep{2011arXiv1104.0223D}. However a similar analysis
at $b=-6^{\circ}$ reveals the presence of two peaks at
$v_r=+100~/-100~$km\,s$^{-1}$ in the faint/bright RC, respectively
(Vasquez et al.  2011, in preparation). Proper motions also do not
seem to differ between stars in the two clumps
\citep{2007AJ....134.1432}.

Further investigations are clearly necessary to completely understand
our bulge structure, kinematics, chemistry and, ultimately,
origin. The present mapping traces the X-shaped structure across the
whole bulge area and may help to guide further analysis, as well as to
design future observations of the Milky Way bulge.

\acknowledgments

We thank the anonymous referee for the useful comments on our
manuscript. We also thank Alvio Renzini for the constructive
discussions while we were preparing this paper. RS acknowledges
financial support from CONICYT through GEMINI Project Nr.  32080016.
MZ and DM are partly supported by Proyectos FONDECYT Regular 1110393
and 1090213, and by Proyecto Conicyt Anillo ACT-86. We gratefully
acknowledge use of data from the ESO Public Survey programme ID
179.B-2002 taken with the VISTA telescope, and data products from the
Cambridge Astronomical Survey Unit, and funding from the FONDAP Center
for Astrophysics 15010003, the BASAL CATA Center for Astrophysics and
Associated Technologies PFB-06, the MILENIO Milky Way Millennium
Nucleus from the Ministry of Economy´s ICM grant P07-021-F, and the
FONDECYT from CONICYT. 2MASS is a joint project of the University of
Massachusetts and the Infrared Processing and Analysis
Center/California Institute of Technology, funded by the National
Aeronautics and Space Administration and the National Science
Foundation.

\end{document}